\documentclass[aps,prl,twocolumn,groupedaddress,floatfix]{revtex4}
\bibpunct{[}{]}{,}{n}{,}{,}
\usepackage{lineno,hyperref}
\usepackage{comment}
\usepackage{graphicx}
\usepackage{dcolumn}
\usepackage{bm}
\usepackage{epsfig}
\usepackage{fancyhdr}
\usepackage{amsmath}
\usepackage{amsfonts}
\usepackage{booktabs}
\usepackage{textcomp}
\usepackage{xcolor}
 \usepackage{siunitx}

\usepackage{float}
\usepackage[T1]{fontenc} 

\begin{document}

\title{Complete set of bound negative-parity states in the neutron-rich $^{18}$N nucleus}

\author{
S. Ziliani$^{1,2}$, M. Ciema\l a$^{3}$, 
F.C.L. Crespi$^{1,2}$, S. Leoni$^{1,2}$\footnote{Corresponding authors: Silvia.Leoni@mi.infn.it,\\Bogdan.Fornal@ifj.edu.pl}, B. Fornal$^{3*}$, 
T. Suzuki$^{4,5}$, T. Otsuka$^{6,7}$, \\
A. Maj$^{3}$,  P. Bednarczyk$^{3}$, G. Benzoni$^{2}$,
A. Bracco$^{1,2}$, C. Boiano$^{2}$, S. Bottoni$^{1,2}$, S. Brambilla$^{2}$, M. Bast$^{8}$, \\
M.  Beckers$^{8}$, T. Braunroth$^{8}$, F. Camera$^{1,2}$,
N. Cieplicka-Ory\'nczak$^{3}$, E. Cl\'ement$^{9}$, S. Coelli$^{2}$, O. Dorvaux$^{10}$, \\
S. Erturk$^{11}$, G. de France$^{9}$, C. Fransen$^{8}$, A.  Goldkuhle$^{8}$,
J. Grebosz$^{3}$, M.N. Harakeh$^{12}$, \L.W. Iskra$^{2,3}$, \\
B. Jacquot$^{9}$, A. Karpov$^{13}$, M. Kici\'nska-Habior$^{14}$, Y. Kim$^{9,15}$, M. Kmiecik$^{3}$,
 A. Lemasson$^{9}$, S.M. Lenzi$^{16,17}$, \\ M. Lewitowicz$^{9}$, 
 H. Li$^{9}$, I. Matea$^{18}$, K. Mazurek$^{3}$, C. Michelagnoli$^{15}$,
 M. Matejska-Minda$^{19,3}$, B. Million$^{2}$, \\
 C. M\"uller-Gatermann$^{8,20}$,  V. Nanal$^{21}$, P. Napiorkowski$^{19}$, D.R. Napoli$^{22}$,
  R. Palit$^{21}$,
   M. Rejmund$^{9}$, Ch. Schmitt$^{10}$, \\ M. Stanoiu$^{23}$, I. Stefan$^{18}$, E. Vardaci$^{24}$ , B. Wasilewska$^{3}$, O. Wieland$^{2}$, M. Zieblinski$^{3}$, M. Zieli\'nska$^{25}$}  

\affiliation{$^1$ Dipartimento di Fisica, Universit$\grave{a}$ degli Studi di Milano, I-20133 Milano, Italy}

\affiliation{$^2$ INFN sezione di Milano, via Celoria 16, 20133, Milano, Italy}

\affiliation{$^3$ Institute of Nuclear Physics, PAN, 31-342 Krak\'ow, Poland}

\affiliation{$^4$ Department of Physics, College of Humanities and Sciences, Nihon University, 3-25-40 Sakurajosui, Setagaya-ku, Tokyo 156-8550, Japan}

\affiliation{$^5$ Division of Science, National Astronomical Observatory of Japan, 2-21-1 Osawa, Mitaka, Tokyo 181-8588, Japan}

\affiliation{$^6$ Department of Physics, The University of Tokyo, 7-3-1 Hongo, Bunkyo, Tokyo 113-0033, Japan}

\affiliation{$^7$ RIKEN Nishina Center, 2-1 Hirosawa, Wako, Saitama 351-0198, Japan}

\affiliation{$^8$ Institut f\"ur Kernphysik, Universit\"at zu K\"oln, 50937 Cologne, Germany}

\affiliation{$^9$ GANIL, CEA/DRF-CNRS/IN2P3, Bd. Henri Becquerel, BP 55027, F-14076 Caen, France}

\affiliation{$^{10}$ Universit\'e de Strasbourg, CNRS, IPHC UMR 7178, F-67000 Strasbourg, France}

\affiliation{$^{11}$ Nigde Omer Halisdemir University, Science and Art Faculty, Department of Physics, Nigde, Turkey}

\affiliation{$^{12}$ KVI - Center for Advanced Radiation Technology, Groningen, Netherlands}

\affiliation{$^{13}$ FLNR, JINR, 141980 Dubna, Russia}

\affiliation{$^{14}$ Faculty of Physics, University of Warsaw, Warsaw, Poland }

\affiliation{$^{15}$ Institut Laue-Langevin (ILL), Grenoble, France}

\affiliation{$^{16}$ INFN Sezione di Padova, I-35131 Padova, Italy}

\affiliation{$^{17}$ Dipartimento di Fisica e Astronomia, Universit$\grave{a}$ degli Studi di Padova, I-35131 Padova, Italy}

\affiliation{$^{18}$ Universit\'e Paris-Saclay, CNRS/IN2P3, IJCLab, 91405 Orsay, France}

\affiliation{$^{19}$ Heavy Ion Laboratory, University of Warsaw, PL 02-093 Warsaw, Poland}

\affiliation{$^{20}$ Physics Division, Argonne National Laboratory, Argonne, Illinois 60439, USA}

\affiliation{$^{21}$ Tata Institute of Fundamental Research, Mumbai 400005, India}

\affiliation{$^{22}$ INFN Laboratori Nazionali di Legnaro, I-35020 Legnaro, Italy}

\affiliation{$^{23}$ IFIN-HH, Bucharest, Romania}

\affiliation{$^{24}$ Universit$\grave{a}$ degli Studi di Napoli and INFN sez. Napoli, Italy}

\affiliation{$^{25}$ IRFU, CEA/DRF, Centre CEA de Saclay, F-91191 Gif-sur-Yvette Cedex, France}

\date{\today}

\begin{abstract}
High-resolution $\gamma$-ray spectroscopy of $^{18}$N is performed with the Advanced GAmma Tracking Array AGATA, following deep-inelastic processes induced by an $^{18}$O beam on a $^{181}$Ta target. Six states are newly identified, which together with the three known excitations exhaust all negative-parity excited states expected in $^{18}$N below the neutron threshold. Spin and parities are proposed for all located states on the basis of decay branchings and comparison with large-scale shell-model calculations performed in the \textit{p-sd} space, with the YSOX interaction. Of particular interest is the location of the 0$^-_1$ and 1$^-_2$ excitations, which provide strong constrains for cross-shell \textit{p-sd} matrix elements based on realistic interactions, and help to simultaneously reproduce the ground and first-excited states in  $^{16}$N and $^{18}$N, for the first time. Understanding the $^{18}$N structure may also have significant impact on neutron-capture cross-section calculations in r-process modeling including light neutron-rich nuclei.
\end{abstract}


\maketitle

The structure of light nuclei can be predicted by state-of-the-art \textit{ab initio} as well as large-scale shell model calculations \citep{Hag14, Heb15, Nav16, Lap16, Kum17, For13, Sax19, Ots10,Yua12,Tsu20,Ots20,Str21}. Both approaches aim at probing nuclear interactions and describing nuclear properties in a wide range of nuclei, including exotic systems, \textit{i.e.}, those lying far away from the stability line.
Of particular interest are  \textit{p-sd} nuclei, for which the neutron dripline has been reached \cite{Tho13,Gui90,Fau96,Tar97,Sak99,Nak17,Ahn19} and their structure also has a significant impact on nuclear astrophysics \cite{Ter01,Her99,Rau94,Bro19,Cow21}.  To reach sufficient accuracy in the description of level ordering, decay branchings, etc., a validation of nuclear structure theory/model predictions is needed. In this context, moderately neutron-rich systems which can be accessed in spectroscopic studies are an ideal testing ground. 

In this work, we focus on the poorly known $^{18}$N nucleus, belonging to the light neutron-rich nuclei 
which are critical for the r-process nucleosynthesis in supernovae \cite{Ter01,Her99,Rau94,Bro19}. $^{18}$N has one proton hole in the \textit{p} shell and three neutrons outside the N = Z = 8 core, \textit{i.e.,} in the \textit{sd} shell. As such, it is a good testing ground for multi-shell \textit{p-sd} interactions, which are employed in large-scale shell-model calculations to reproduce simultaneously the structural properties of neutron-rich light nuclei and their drip lines. For example, the YSOX interaction of Ref. \cite{Yua12}, in which the cross-shell $<$$psd|V|psd$$>$ and $<pp|V|sdsd>$ matrix elements are based on the monopole-based universal interaction $V_{MU}$, while the intershell matrix elements are phenomenological, has been successful in reproducing ground-state energies, driplines and energy levels of most of
\textit{p-sd} shell nuclei. In particular, this interaction correctly predicted the ordering of low-lying states in N = 11 isotones, including $^{18}$N, where other interactions, such as WBP and WBT \cite{War92}, fail. On the contrary, shell model calculations with YSOX did not solve the long-time problem of simultaneously explaining the level ordering in $^{16}$N and $^{18}$N \cite{Yua12}. 

In this paper, we aim at a complete $\gamma$ spectroscopy of bound states in $^{18}$N, which will be instrumental in constraining cross-shell interactions \cite{Tsu14} in this region of the nuclear chart. We used deep-inelastic processes, induced by an $^{18}$O beam on a thick $^{181}$Ta target \cite{Cie20,Cie21}, to populate states in $^{18}$N, and the Advanced GAmma Tracking Array (AGATA) \cite{Akk12,Cle17,Kor20}, to detect the $\gamma$ decays from these states. Based on the analysis of the collected data, we propose the location of all negative-parity states, below the neutron-separation energy, including 0$^{-}$ and 1$^{-}$ states arising from the coupling between a proton hole in the $p_{1/2}$ and a neutron in the $s_{1/2}$ orbitals. Such data allow to investigate further details of the proton-neutron interaction, including specific proton-neutron matrix elements which play a key role in the level ordering of both $^{18}$N and $^{16}$N nuclei. In addition, a firm location of the first-excited 1$^{-}$ state of $^{18}$N is of high importance for the calculated neutron-capture reaction rate on $^{17}$N --  it can change this cross section by up to one order of magnitude \cite{Her99,Rau94}. 

Previous investigations of $^{18}$N, performed by employing charge-exchange \cite{Put83}, fusion evaporation \cite{Wie08}, (d,p) \cite{Hof13} reactions and beta-decay \cite{Pra91}, located a total of 8 levels above the 1$^-_1$ ground state (g.s.). Three excited states were firmly identified below 1 MeV, at 115, 587 and 742 keV, with spin 2$^-_1$, (2$^-_2$) and 3$^-_1$, respectively. The gamma decay between them was also observed \cite{Wie08,Pra91,NUDAT}. At higher energies, three additional levels were placed at 1.17(2), 2.21(1) and 2.42(1) MeV, with large energy uncertainties. The level at 1.17(2) MeV was observed in the (d,p) reaction study of Hoffman \textit{et al.}, \cite{Hof13}, and interpreted as a (1$^-_2$) state, or a doublet of unresolved 0$^-_1$ and 1$^-_2$ states. Finally, two levels located at 1735 and 2614 keV, in beta-decay studies, were tentatively proposed as positive-parity (1$^+$) states \cite{Pra91}.

Within the shell-model framework, a limited number of states is expected in $^{18}$N below the neutron separation energy, S$_n$ = 2.828(24) MeV. In particular, calculations performed with different interactions \cite{Yua12} predict 10 negative-parity states with spin-parity 1$^-$ (two states, including the 1$^-_1$ g.s.), 2$^-$ (three states), 3$^-$ (three states), 0$^-$ (one state) and 4$^-$ (one state). These lowest-lying states can be interpreted as arising from the coupling of a proton in the $p_{1/2}$ orbital to the lowest members of the multiplet of states originating from \textit{i)} three neutrons in the $d_{5/2}$ orbital or \textit{ii)} two neutrons (coupled to spin 0) in the $d_{5/2}$ and one neutron in the $s_{1/2}$ orbitals. Along this line, the 1$^-_1$ g.s. and the first three excited states of $^{18}$N can be viewed, for example, as the coupling between a proton in the $p_{1/2}$ orbital with the 3/2$^+$ g.s. and the 5/2$^+$ first-excited state in $^{17}$C, producing the doublets (1$^-_1$ g.s., 2$^-_2$) and (2$^-_1$, 3$^-_1$), respectively. Other negative-parity states, which are expected below $S_n$ in $^{18}$N, should arise from the coupling of the $p_{1/2}$ proton with higher-lying states in $^{17}$C, which are unbound and not fully known experimentally (in $^{17}$C, S$_n$ = 734 keV). Positive-parity states in $^{18}$N are instead calculated at higher energies, above S$_n$. In this work, we perform a search for all bound states in $^{18}$N by employing a non-selective reaction mechanism and an efficient experimental setup, with state-of-the-art energy resolution for gamma detection.

\begin{figure}[ht]
\centering
\resizebox{0.48\textwidth}{!}{\includegraphics{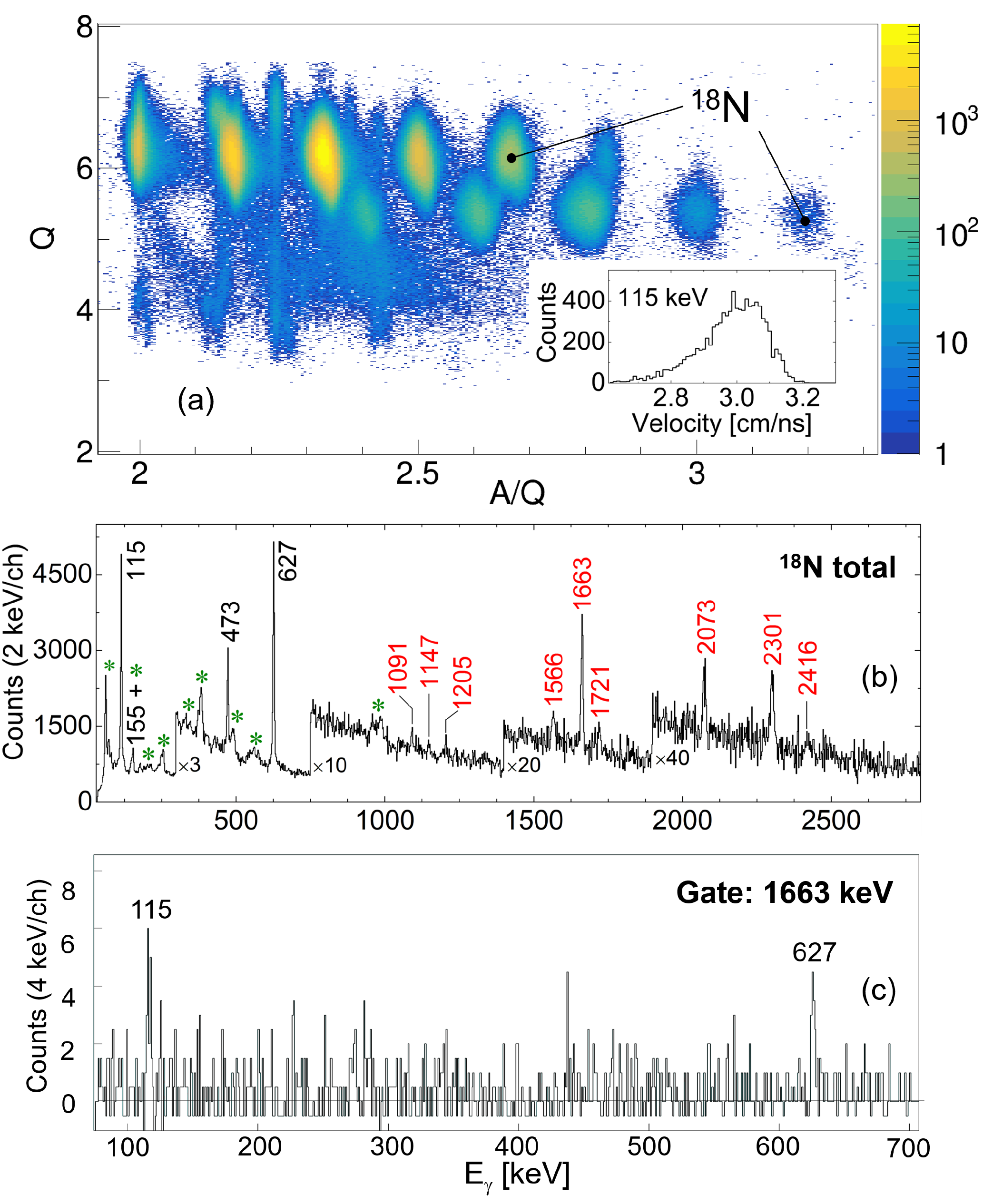}}
\caption{(a): identification plots, ion charge Q versus ratio of mass number A to Q, for N ions, as measured by VAMOS++. Inset: velocity distribution measured at the VAMOS++ focal plane, in coincidence with the 2$^-_1 \rightarrow$ 1$^-_1$, 115-keV, ground-state transition of $^{18}$N. (b):  Doppler-corrected $\gamma$-energy spectrum of $^{18}$N, as measured by the AGATA array. Energies of already known transitions are shown in black \cite{Pra91,Wie08,NUDAT}, those of newly observed gamma rays in red. Contaminant lines from $^{180,181}$W binary-reaction partners are marked by green stars. (c): $\gamma$-ray spectrum in coincidence with the 1663-keV line, showing the coincidences with the 115- and 627-keV transitions.} 
\label{spectra}
\end{figure}

In our experiment, $^{18}$N was populated in deep inelastic processes \cite{Sch85,Zag14,Kar17} induced by an $^{18}$O beam at 126 MeV (7.0 MeV/u) on a $^{181}$Ta target, 4 $\mu$m thick (6.64 mg/cm$^2$). At the target center, the beam energy was $\sim$116 MeV (i.e., 50$\%$ above the Coulomb barrier), leading to projectile-like products with v/c$\sim$10$\%$ relative velocity. The experiment was performed at the Grand Accélérateur National d’Ions Lourds (GANIL) with the $\gamma$-tracking array AGATA \cite{Akk12,Cle17,Kor20} (consisting of 31 high-purity Ge detectors) coupled to a scintillator array made of two large-volume (3.5"$\times$8") LaBr$_3$ detectors plus two clusters of the PARIS setup \cite{Maj09}.  The reaction products were detected in the VAMOS++ magnetic spectrometer \cite{Rej11} placed at the reaction grazing angle (\textit{i.e.}, 45$^{\circ}$ $\pm$ 6$^{\circ}$ with respect to the beam direction), and aligned with the center of AGATA.  The tracking array covered the 115$^{\circ}$-175$^{\circ}$ backward angular range, with respect to the VAMOS++ axis, while the scintillators were placed at 90$^{\circ}$. A total of more than 10$^7$ events were collected requiring the projectile-like products detected in VAMOS++, in coincidence with gamma rays in AGATA or PARIS. The VAMOS++ spectrometer setting, optimized to detect $^{20}$O \cite{Cie20} within a large velocity range, allowed to detect other products with charge 5$\leq Z \leq$9 and mass number 11$\leq A \leq$21. Figure \ref{spectra} (a) shows the identification plot of the nitrogen ions, corresponding to a total of 3.8 $\times$ 10$^6$ events, out of which 2.6 $\times$10$^5$ belonging to  $^{18}$N. The inset gives the velocity distribution of  $^{18}$N associated with the population of the first-excited 2$^-_1$ state at 115 keV: it displays a significant tail towards lower velocity, supporting the production of  $^{18}$N in deep-inelastic processes \cite{Sch85,Zag14,Kar17}.

\begin{figure*}[ht]
\centering
\resizebox{1.0\textwidth}{!}{\includegraphics{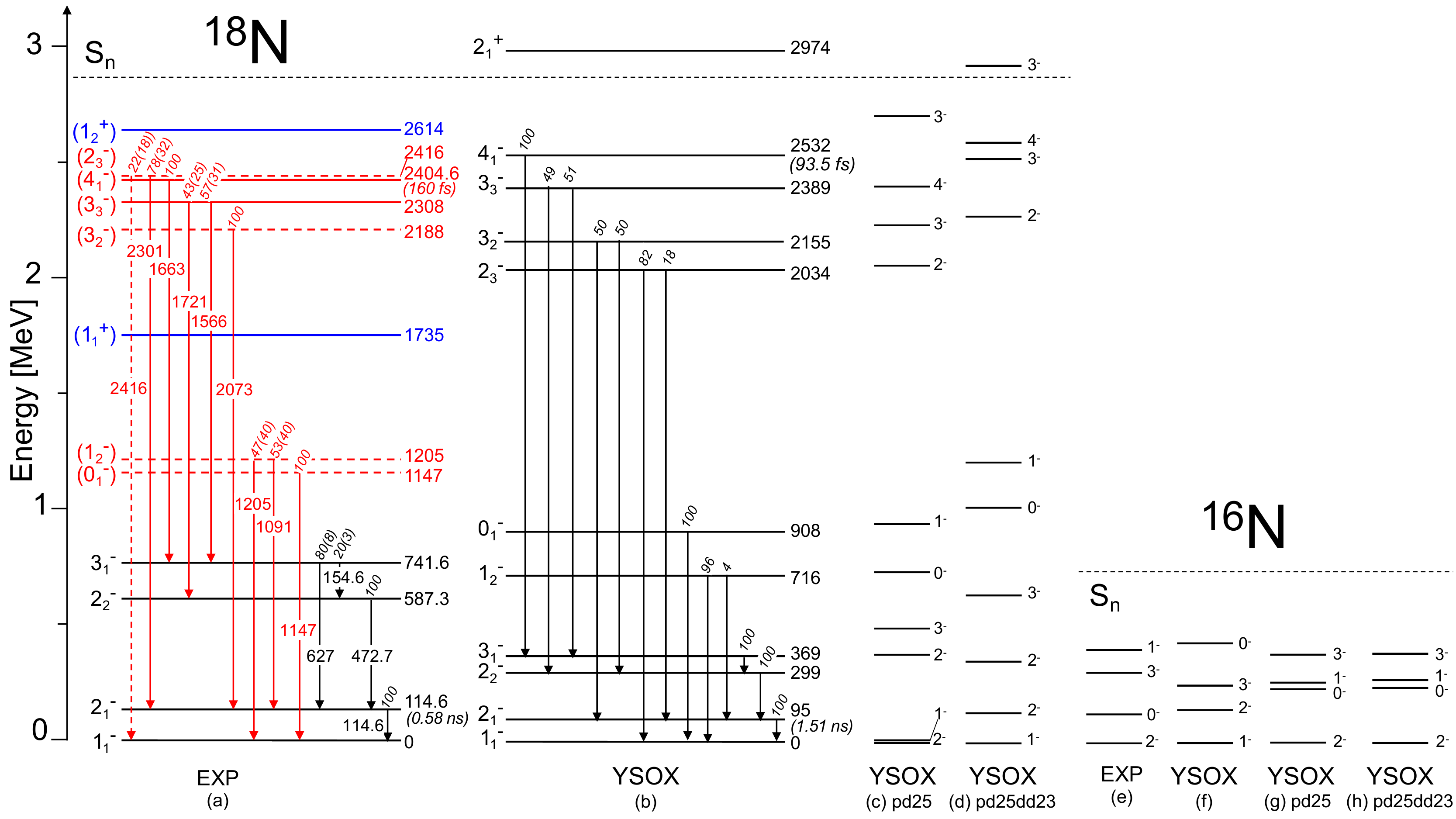}}
\caption{(a): experimental decay scheme of $^{18}$N, as obtained in the present AGATA experiment. In black, previously-known levels and transitions \cite{Pra91,Wie08,NUDAT}; in red, newly found ones (dashed lines for tentative). Positive-parity states observed in beta-decay studies \cite{Pra91} are marked in blue.  (b): decay schemes of $^{18}$N predicted by shell-model calculations using the YSOX interaction of Ref. \cite{Yua12}. (c)-(d): $^{18}$N shell-model predictions with YSOX, with modified matrix elements involving proton-$p_{1/2}$-neutron-$s_{1/2}$ (c) and \textit{sd} (d) orbitals. (e): experimental bound states of $^{16}$N \cite{NUDAT}. (f)-(h): shell-model predictions of $^{16}$N obtained with the original \cite{Yua12} (f) and modified YSOX interactions (g)-(h), as in panels (c) and (d) (see text for details).} 
\label{levels}
\end{figure*}

\begin{figure}[ht]
\centering
\resizebox{0.48\textwidth}{!}{\includegraphics{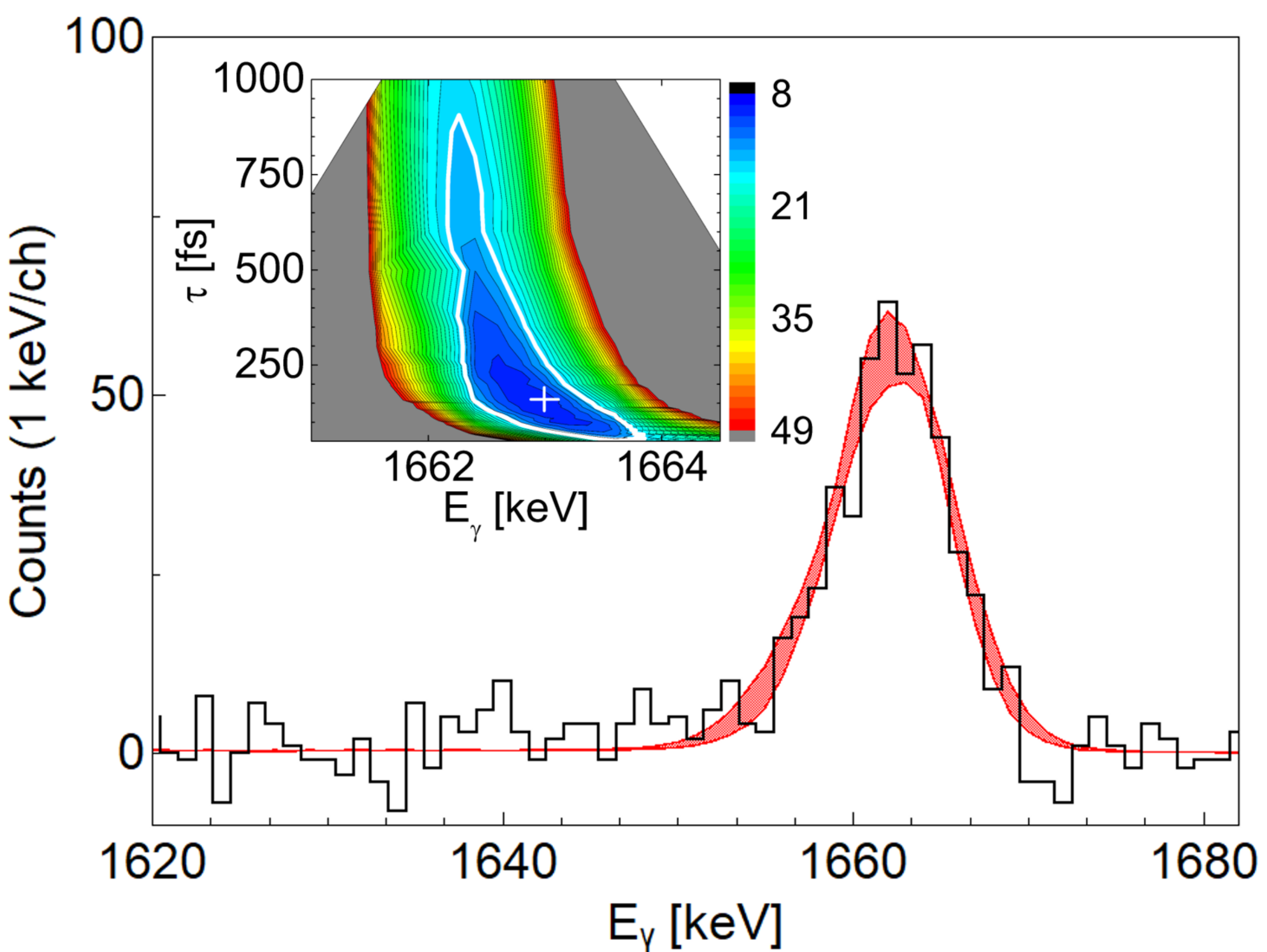}}
\caption{Gamma-ray energy spectrum of $^{18}$N in the region of the 1663-keV $\gamma$ ray, as measured with AGATA over the entire angular range (black histogram).  The red-shaded band is the result of the lineshape simulation with the method of Refs. \cite{Cie20,Cie21}, performed by varying E$_{\gamma}$ and $\tau$ within the uncertainty region of the corresponding two-dimensional $\chi^2$ lifetime-energy surface shown in the inset (the white cross and white contour line indicate the minimum and the uncertainty region, corresponding
to 80$\%$ confidence level), e.g., E$_{\gamma}$ = 1663.0(8) keV and $\tau$ = 160$^{+740}_{-100}$ fs. } 
\label{lifetime}
\end{figure}

The $\gamma$-ray spectrum obtained with AGATA by gating on $^{18}$N ions, Doppler-corrected event-by-event by considering the product velocity vector measured in VAMOS++, is shown in Fig. 1 (b) \cite{Cie21}.  Previously known 114.6-, 155-, 472.7- and 627-keV transitions are marked in black \cite{Pra91,Wie08,NUDAT}, while newly observed $\gamma$ rays at 1091(1), 1147(1), 1205(1), 1566.0(10), 1663.0(8), 1721.0(11), 2073.4(8), 2301.0(8), and 2416(2) keV are indicated by red labels. Green stars mark lines from $^{180,181}$W binary-reaction partners, which are broadened and shifted due the Doppler-shift correction for the $^{18}$N product, applied to this spectrum. Spectra constructed by gating on the 114.6-keV (2$^-_1$→ 1$^-_1$) and 627-keV (3$^-_1$→ 2$^-_1$) transitions confirmed the coincidence relationships between the 114.6-, 472.7-, 155- and 627-keV $\gamma$ rays, reported earlier \cite{Wie08}. They also showed the presence of the newly found 1663.0-keV line (see Fig. \ref{spectra}(c)). Such a transition was therefore placed in cascade with the 627.0-keV and 114.6-keV gamma rays, depopulating a state at 2404.6 keV, as shown in Fig. \ref{levels} (a). None of the other new lines could be seen in the coincidence spectra, either due to the limited statistics or their possible direct feeding to the 1$^-$ ground state. Next, by inspecting energy differences between gamma rays, three new levels were identified. First, we considered the 1566.0- and 1721.0-keV transitions which differ by 155 keV, what equals the energy difference between the 3$^-_1$ and 2$^-_2$ states, at 741.6 and 587.3 keV. One may then assume that they deexcite a state at E$_{exc}$ = 2308 keV, with relative branchings of 57(31) and 43(25), respectively. Similarly, the 2301- and the weak 2416-keV gamma rays could deexcite a level placed at 2416 keV, feeding the first-excited 2$^-_1$ and the 1$^-_1$ ground states, with relative branchings of 78(32) and 22(18), respectively. The group of three weak transitions observed at 1091, 1147 and 1205 keV required special attention, as they could be related to the existence of a 1$^-$ state or a doublet of unresolved 0$^-_1$ and 1$^-_2$ states, reported by Hoffman \textit{et al. }\cite{Hof13} at 1.17(2) MeV. It is very likely that the 1205- and 1091-keV gamma rays, having an energy difference of $\sim$115 keV, feed the 1$^-_1$ g.s. and the 2$^-_1$ state (at 115 keV) from a level at E$_{exc}$ = 1205 keV, with relative branchings of 47(40) and 53(40), respectively. The other 1147-keV line could instead populate directly the 1$^-_1$ g.s. from a level at E$_{exc}$ = 1147 keV. This placement is consistent with the existence of a doublet of unresolved states with 1.17-MeV average energy. Finally, the remaining newly observed 2073-keV line was proposed to feed the 2$^-_1$ state from a level at 2188 keV, in agreement with both the indication of a presence of a state at $\sim$2.2 MeV, as reported in charge-exchange and (d,p) studies \cite{Pra91, Hof13}, and theory considerations discussed below. The resulting level scheme is presented on the left of Fig. \ref{levels} (a). In this figure, lifetime information is also given for the 114.6-keV state ($\tau$ = 0.58(16) ns, from Wiedeking \textit{et al.} \cite{Wie08}), and for the state at 2404.6 keV. In the latter case, the lifetime value $\tau$ = 160$^{+740}_{-100}$ fs was extracted from a lineshape analysis of the intense 1663-keV transition, by employing the Monte Carlo procedure described in Refs. \cite{Cie20,Cie21}. As shown in Fig. \ref{lifetime}, the procedure also yielded the most precise transition energy value E$_{\gamma}$ = 1663.0(8) keV.  

In previous works, the structure of the bound states in $^{18}$N was calculated by using a shell-model approach with various interactions: the WBP and WBT interactions \cite{War92}, and the YSOX interaction \cite{Yua12} mentioned above. In this work, we interpret our experimental findings with the help of the YSOX shell-model calculations. In Fig. 2 (b), the $^{18}$N level and decay scheme calculated with the YSOX interaction are displayed. As discussed in earlier works, the ordering of the first four states (1$^-_1$, 2$^-_1$, 2$^-_2$ and 3$^-_1$) is well reproduced (contrary to the case of WBP and WBT), as well as the decay pattern. However, the second- and third-excited states are predicted at lower energies, approximately at half the excitation energies observed in the experiment. A few hundreds keV above the third excited state, 1$^-$ and 0$^-$ excitations are predicted, lying $\sim$200 keV apart from each other. They correspond to the experimental doublet located at 1147 and 1205 keV. On the basis of the comparison between calculated and observed decay patterns (\textit{i.e.}, two branches from the level at 1205 keV to the 2$^-_1$ and 1$^-_1$ g.s., and a decay from the level at 1147 keV exclusively to the 1$^-_1$ g.s.), we tentatively assign spin-parity 0$^-$ to the 1147- and 1$^-$ to the 1205-keV states. Above 2 MeV, calculations predict five negative-parity states. They can be related to the five states located experimentally. The level at 2405 keV is the most strongly populated. Based on the fact that deep-inelastic reactions populate preferentially yrast states, we assign to it spin-parity 4$^-$. In this case, the experimental lifetime value could be determined (see Fig. \ref{lifetime}), yielding the value $\tau$ = 160$^{+740}_{-100}$ fs, which is in line with the calculated value of 93.5 fs. Further, the decay pattern of the experimental 2308- and 2416-keV levels are consistent with the calculated decay schemes of the 3$^-_3$ and 2$^-_3$ states at 2389 and 2034 keV. Therefore, we assign to them spin-parity 3$^-$ and 2$^-$, respectively. Consequently, the level at 2188 keV will have spin-parity assignment of 3$^-$, as it is associated with the calculated 3$^-_2$ state at 2155 keV.

As shown in Fig. \ref{levels}(a) and (b), the shell-model predictions with the YSOX interaction, although providing a rather satisfactory description of the excitation spectrum and decay scheme of $^{18}$N, do not
reproduce the ordering, in particular, of the 0$^-_1$ and 1$^-_2$ states around 1.17 MeV – these states arise from the coupling between a proton hole in the $p_{1/2}$ and a neutron in the $s_{1/2}$ orbitals. The ordering of higher-lying states, above 2 MeV, is more difficult to interpret, since the location of such states may be affected by being in the vicinity of the neutron threshold. 

An attempt was made to improve the agreement between data and shell-model predictions, by adjusting selected cross-shell \textit{p-sd} two-body matrix elements by about 20-30$\%$. In particular, to reverse the order of the 0$_{1}^{-}$ and 1$_{2}^{-}$ levels in $^{18}$N and become consistent with the experimental data, the matrix elements for the $\pi p_{1/2} -\nu s_{1/2}$ orbits were adjusted ($<$$\pi p_{1/2}, \nu s_{1/2}(J)|V|\pi p_{1/2}, \nu s_{1/2}(J)$$>$ were varied by -0.3 (+0.2) MeV for J$^{\pi}$ = 0$^{-}$ (1$^{-}$)), while the spacing among calculated levels was improved by varying the matrix element involving the $\pi p_{1/2}$--$\nu d_{5/2}$ orbits ($<$$\pi p_{1/2}, \nu d_{5/2}(J)|V|\pi p_{1/2},\nu d_{5/2}(J)$$>$ was modified by -0.25 MeV for J = 2$^{-}$). 
A modification of two-body matrix elements within the \textit{sd}-shell part ($<$$\nu d_{5/2}^2(J)|V|\nu d_{5/2}^2 (J)$$>$ was varied by -0.23 MeV for J = 2$^{+}$) widened the energy gaps among the (1$^{-}_1$ g.s., 2$^{-}_2$), (2$^{-}_1$, 3$^{-}_1$) and (0$^{-}_1$, 1$^{-}_2$) pairs. This change lowers the $d_{5/2}$-orbit, and decreases the energies of the pairs (1$^{-}$, 2$^{-}$), (2$^{-}$, 3$^{-}$), whose main components are $\pi p_{1/2}\nu d_{5/2}^3$ configurations, relative to the pair (0$^{-}$, 1$^{-}$), that is dominantly of $\pi p_{1/2}\nu d_{5/2}^2 s_{1/2}$ origin.  As a result, the gap between the four lowest lying states and the (0$^{-}$, 1$^{-}$) pair is enhanced (Fig. \ref{levels} (d)). Moreover, an improvement in the transition probabilities is obtained. In particular, for the decay from the 3$^-_1$ state, two branches leading to 2$^-_1$ and 2$^-_2$ with similar intensities are calculated, in better agreement with the experiment.

The changes of matrix elements introduced above lead also to the correct reproduction of the spin and parity of the ground state and first-excited state in $^{16}$N, as shown in Fig. \ref{levels}(e), (g) and (h). The 2$^{-}$ level in $^{16}$N, which is dominantly of $\pi p_{1/2}\nu d_{5/2}$ nature, is lowered by the more attractive $\pi p_{1/2}$--$\nu d_{5/2}$ interaction, and becomes the ground state in agreement with the experiment. The 0$^-$ excitation becomes the first-excited state, after the adjustment of the $\pi p_{1/2}$--$\nu s_{1/2}$ cross-shell matrix elements, which reverses the order of the 0$^-_1$ and 1$^-_1$ states. 

Altogether, the improved description of the $^{18}$N and $^{16}$N data appears to be related to the adjustment of the monopole term for $\pi p_{1/2} - \nu d_{5/2}$ ,which makes it more attractive than in the original YSOX interaction \cite{Yua12}. Some modifications in the multipole terms are also necessary for $\pi p_{1/2} - \nu s_{1/2}$ and $\pi p_{1/2} -\nu d_{5/2}$ to obtain a better agreement with the experimental data.  We note that the dripline of the N isotopes remains at $^{23}$N for both the modified YSOX interactions.

In summary, in this work we have performed a high-resolution $\gamma$-spectroscopy investigation, with the AGATA array, of the $^{18}$N nucleus populated in deep-inelastic processes induced by an $^{18}$O beam on a $^{181}$Ta target. A total of six states have been newly identified, which together with three already known excitations exhaust all negative-parity excited states expected in $^{18}$N below the neutron threshold. Large-scale shell-model calculations performed in the \textit{p-sd} space, with the YSOX interaction,  reasonably reproduce the experimental data, apart from the ordering of the 0$^-_1$ and 1$^-_2$ states, which originate from the coupling between a $p_{1/2}$ proton and a $s_{1/2}$ neutron. Selective variations of two-body \textit{p-sd} cross-shell and \textit{sd} inter-shell matrix elements restore the level ordering in $^{18}$N, simultaneously reproducing the ground and first-excited state in $^{16}$N, for the first time.  These results help constraining cross-shell proton-neutron effective interactions in the \textit{p-sd} space, so far little explored in comparison with similar investigations in the \textit{sd-pf} shell \cite{Tsu17}.  Of particular interest is also the location of the 1$^-_2$ excitation in $^{18}$N, which has strong impact on neutron-capture cross-section calculations in r-process modeling including light neutron-rich nuclei \cite{Ter01,Her99,Rau94}. 

The work demonstrates the power of state-of-the-art instruments, such as the AGATA \cite{Akk12,Cle17,Kor20} and GRETINA \cite{Pas13,Fal16} tracking arrays and complementary detectors, in performing complete spectroscopy of hard-to-reach nuclei lying on the neutron-rich side of the stability valley.

\bigskip

The authors wish to thank the staff of the GANIL Laboratory for providing the beam and for the help given during the experiment. The AGATA collaboration is also acknowledged. This work was supported by the Italian Istituto Nazionale di Fisica Nucleare, by the Polish National Science Centre under Contracts No. 2014/14/M/ST2/00738, No. 2013/08/M/ST2/00257, No. 2016/22/M/ST2/00269, and the NCN project No. 2014/12/S/ST2/00483, and by the RSF Grant No. 19-42-02014 and by the U.S. Department of Energy, Office of Science, Office of Nuclear Physics, under contract number DE AC02-06CH11357. This project has received funding from the Turkish Scientific and Research Council (Project No. 115F103) and from the European Union Horizon 2020 Research and Innovation Program under Grant Agreement No. 654002. This work was supported in part by MEXT KAKENHI Grant Nos. JP19H05145 and JP19K03855.

\bibliography{Letter-PRC-18N-SUBMITTED}
\newpage

\end{document}